\documentclass[epjCONF]{svjour}
\usepackage{graphics}
\usepackage[varg]{txfonts} 
\usepackage[latin1]{inputenc}
\session-title{UHECR 2012 Symposium}
\begin{document}
\title{Ultra High Energy Particles Propagation and the Transition from Galactic to Extra-Galactic Cosmic Rays}
\author{Roberto Aloisio\inst{1,2}\fnmsep\thanks{\email{aloisio@arcetri.astro.it}}  }
\institute{INAF - Osservatorio Astrofisico di Arcetri, Firenze, Italy.  \and 
INFN - Laboratori Nazionali del Gran Sasso, L'Aquila, Italy.}
\abstract{We discuss the basic features of the propagation of Ultra High Energy Cosmic Rays in astrophysical 
backgrounds, comparing two alternative computation schemes to compute the expected fluxes. We also discuss 
the issue of the transition among galactic and extra-galactic cosmic rays using theoretical results on fluxes 
to compare different models.
} 
\maketitle
\section{Introduction}
\label{intro}

Ultra High Energy Cosmic Rays (UHECR) are the most energetic particles observed 
in nature with (detected) energies up to $3\div 5 \times 10^{20}$ eV. The experimental 
observations of UHECR are performed nowadays by the Auger observatory in Argentina 
\cite{Auger} and by the HiRes \cite{HiRes} and Telescope Array (TA) \cite{TA} observatories 
in the USA. The observation of these particles poses many interesting questions mainly on 
their nature and origin. 

The study of the propagation of UHECR through astrophysical backgrounds can be very useful to 
interpret the observations and, may be, to have important insights on the possible sources of these 
particles. The propagation of UHECR from the sources to the observer is 
mainly conditioned by the interactions with the intervening astrophysical 
backgrounds such as the Cosmic Microwave Background (CMB) and the Extragalactic 
Background Light (EBL). While the propagation of protons is conditioned only by the CMB 
radiation field \cite{Nuclei}, nuclei propagation is also affected by the EBL radiation \cite{Nuclei}. 

Several propagation dependent features in the spectrum can be directly linked to the chemical 
composition of UHECR and/or to the distribution of their sources \cite{Boncioli}. Among such 
features particularly important is the Greisin, Zatsepin and Kuzmin (GZK) suppression of the flux, 
an abrupt depletion of the observed proton spectrum due to the interaction of the UHE protons 
with the CMB radiation field \cite{GZK}. The GZK suppression, as follows from the original papers, 
is referred to protons and it is due to the photo-pion production process on the CMB radiation 
field ($p+\gamma_{CMB} \to \pi + p$). In the case of nuclei the expected flux also shows a 
suppression at the highest energies that, depending on the nuclei specie, is due to the 
photo-disintegration process on the CMB and EBL radiation fields 
($A + \gamma_{CMB,EBL} \to (A - nN) + nN$) \cite{Nuclei}. In any case, the interaction processes 
between UHE particles and astrophysical backgrounds will condition the end of the CR 
spectrum at the highest energies and the high energy behavior of the flux can be used 
as a diagnostic tool for the chemical composition of the observed particles. 
Another important feature in the spectrum that can be directly linked with the nature of the 
primary particles and their origin (galactic/extra-galactic) is the pair-production dip \cite{dip}. 
This feature is present only in the spectrum of UHE extragalactic protons and, as the 
GZK, is a direct consequence of the proton interaction with the CMB radiation field, in 
particular the dip brings a direct imprint of the pair production process 
$p+\gamma_{CMB} \to p+e^{+} + e^{-}$ suffered by protons in their interaction with CMB 
radiation. 

The experimental observations show contradictory results mainly on the chemical composition of UHECR. 
While Auger favors a light (proton) composition at low energies and an heavy (nuclei) composition at the 
highest energies \cite{Chem}, HiRes and TA show a proton dominated spectrum at all energies \cite{Chem}. 
A clear understanding of the chemical composition of UHECR is of paramount importance in the study of such 
particles in particular in the determination of their possible sources and to tag the transition among galactic and 
extra-galactic CR \cite{TransReview}.  

Apart from the observations on chemical composition, also the observations on fluxes show a certain level 
of disagreement among HiRes and TA on one side and Auger on the other side. HiRes and TA experiments 
show a flux with a clear observation of the protons GZK suppression and the pair-production dip, coherently 
with their composition observations \cite{Flux}. The situation changes if the Auger results are taken into account. 
The latest release of the Auger data on flux shows a spectrum not compatible with the pair production dip with 
an high energy suppression not compatible with the GZK suppression as expected for protons. Signaling a 
possible deviation from a proton dominated spectrum to an heavier composition at the highest energies, 
confirming with the flux behavior the observations on chemical composition \cite{Flux}. 

This puzzling situation, with different experiments favoring different scenarios, shows the importance of a 
systematic study of the propagation of UHECR in astrophysical backgrounds. In the present paper we will 
present two alternative ways to study the propagation of UHECR in astrophysical backgrounds: through a 
Monte Carlo (MC) approach and through an analytic computation scheme based on the kinetic theory of 
particles propagation. We will than compare theoretical results with observations, considering in particular 
the issue of the transition from galactic to extra-galactic CR. 

The paper is organized as follows: in section \ref{prop} we will briefly review theoretical models to study 
UHECR propagation, in section \ref{trans} we will use the results of the propagation models to determine
the features of the transition among galactic and extra-galactic CR, in this section we will use the results of 
\cite{AmatoBlasi} for the galactic component, finally we will conclude in section \ref{conclude}.

\section{Ultra High Energy Cosmic Rays Propagation}
\label{prop}

The propagation of charged particles (protons or nuclei) with energies above $10^{17}$ eV through 
astrophysical backgrounds can be suitably studied taking into account the main channels of interaction 
that, as already anticipated in the introduction, are:

\begin{itemize}
\item{{\it protons}} UHE protons interact only with the CMB radiation field giving rise to the two processes 
of pair production and photo-pion production.
\item{{\it nuclei}} UHE nuclei interact with the CMB and EBL radiation fields, suffering the process of pair 
production, in which only CMB is relevant, and photo-disintegration, that involves both backgrounds. 
While the first process conserves the nuclear species, the second produces a change in the nuclear species, 
extracting nucleons from the nucleus.
\end{itemize}

In the energy range $E\simeq 10^{18} \div 10^{19}$ eV the propagation of UHE particles is extended 
over cosmological distances with a typical path length of the order of Gpc. Therefore we should also 
take into account the adiabatic energy losses suffered by particles because of the cosmological expansion 
of the Universe.

In this section we will briefly recall two alternative computation schemes used to describe the propagation of 
UHECR: the kinetic approach \cite{Nuclei} and the SimProp MC approach \cite{SimProp}.

The kinetic approach is based on the hypothesis of continuum energy losses (CEL), through which particles 
interactions are treated as a continuum process that continuously depletes the particles energy. In the propagation
through astrophysical backgrounds the interaction of particles are naturally affected by fluctuations, with a non-zero 
probability for a particle to travel without loosing energy. The effect of the fluctuations in the interaction of particles 
is taken into account in the MC approach while it is neglected in the case of the kinetic approach (CEL approximation). 
In the case of proton propagation the CEL approximation has a negligible effect on the pair-production process, 
while in the case of photo-pion production it gives a deviation only at the highest energies ($E\ge10^{20}$ eV) of 
the order of $10\%$ with respect to the flux computed taking into account the intrinsic stochasticity of the process 
\cite{dip,SimProp}.
 
\begin{figure}[!htb]
\centering
\begin{tabular}{ll}
\resizebox{0.4\columnwidth}{!}{
\includegraphics{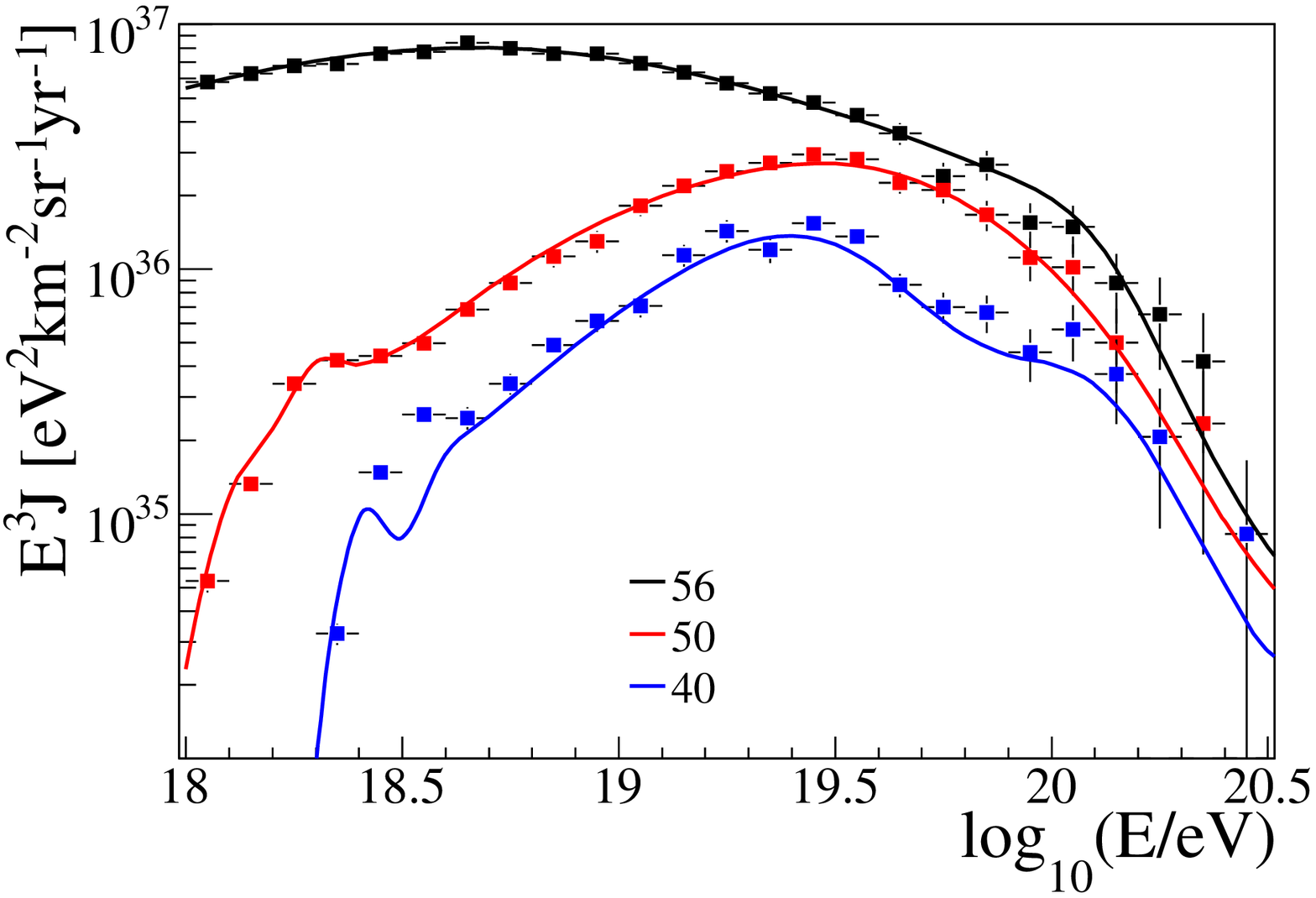} }
&
\resizebox{0.4\columnwidth}{!}{
\includegraphics{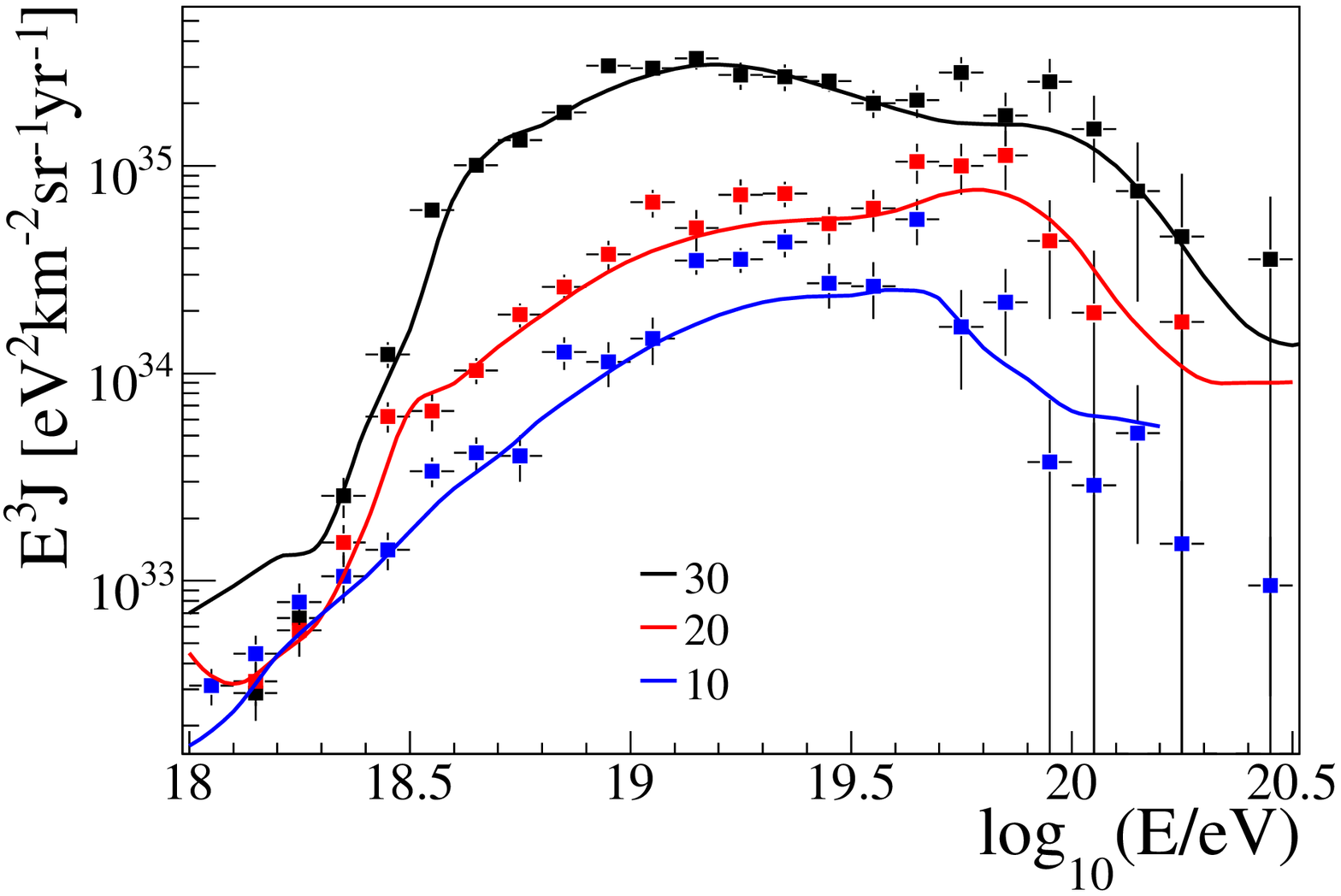} }
\end{tabular}
\caption {\label{fig1}{\small{Flux of iron and secondary nuclei at $z=0$ in the case of pure iron injection at 
the source with a power law injection index $\gamma_g=2.2$. Full squares correspond to the {\it SimProp} result
while continuous lines correspond to the solution of the nuclei kinetic equation. 
Left panel: fluxes of iron and of secondary nuclei with $A=50,40$. 
Right panel: fluxes of secondary nuclei with $A=30,20,10$. }}} 
\end{figure}
 
In the case of nuclei propagation the energy losses due to the process of pair production can be simply related to the 
corresponding quantity for protons \cite{Nuclei}.Therefore the CEL approximation can be suitably used also for 
the pair production process involving nuclei.  

Let us now discuss the process of photo-disintegration of nuclei: this interaction changes the nucleus 
kind leaving its Lorentz factor unchanged. In the kinetic approach of \cite{Nuclei} the process of 
photo-disintegration is treated as a decay process that simply depletes the flux of the nucleus $A$. 
The photo-disintegration "life-time" is defined as:
\begin{equation}
\frac{1}{ \tau_{A}(\Gamma)}= \frac{c}{2 \Gamma^2} \sum_i \int^{\infty}_{\epsilon_{0}(A)} d\epsilon'
\sigma_{A,i}(\epsilon') \nu_i(\epsilon') \epsilon' 
\int^{\infty}_{\epsilon'/(2\Gamma)} d\epsilon \frac{n_{\gamma}(\epsilon)}{\epsilon^2}
\label{eq:betadisi} 
\end{equation}
with $A$ the atomic mass number and $\Gamma$ the Lorentz factor of the interacting particle, $\epsilon'$ 
the energy of the background photon in the rest frame of the particle, $\epsilon_{0}(A)$ the threshold of the 
considered reaction in the rest frame of the nucleus $A$, $\sigma$ the relative cross section, $\nu$ the 
photo-disintegration multiplicity (number of emitted nucleons), $\epsilon$ the energy of the photon in 
the laboratory system and $n_{\gamma}(\epsilon)$ the density of the background photons per unit energy.

The kinetic approach is based on the transport equations that describe the propagation of protons and nuclei 
through astrophysical backgrounds, these equations for protons and nuclei are respectively: 

\begin{equation}
\frac{\partial n_p(\Gamma,t)}{\partial t} - \frac{\partial}{\partial 
\Gamma} \left [ b_p(\Gamma,t)n_p(\Gamma,t) \right ] = Q_p(\Gamma,t)
\label{eq:kin_p}
\end{equation}   
\begin{equation}
\frac{\partial n_{A}(\Gamma,t)}{\partial t} - \frac{\partial}{\partial \Gamma}
\left [ n_{A}(\Gamma,t) b_{A}(\Gamma,t) \right ] + \frac{n_{A}(\Gamma,t)}
{\tau_{A}(\Gamma,t)}  = Q_{A}(\Gamma,t)
\label{eq:kin_A}
\end{equation}
where $n$ is the equilibrium distribution of particles, $b$ are the energy losses (adiabatic expansion of the Universe
and pair/photo-pion production for protons or only pair-production for nuclei) $Q$ is the injection of freshly
accelerated particles and, in the case of nuclei, also the injection of secondary particles produced by 
photo-disintegration (secondary nuclei). 

\begin{figure}[!htb]
\centering
\begin{tabular}{ll}
\resizebox{0.4\columnwidth}{!}{
\includegraphics{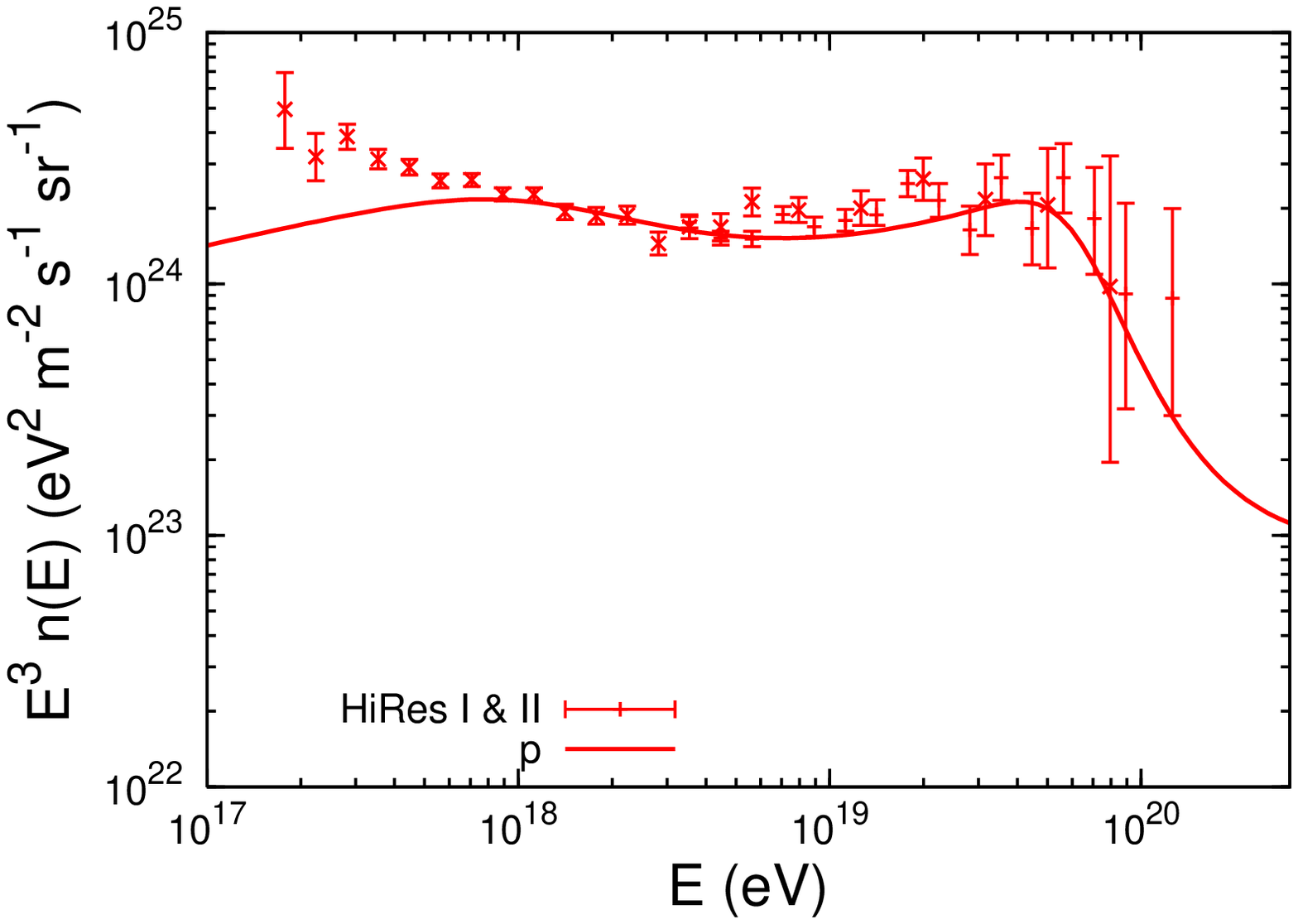} }
&
\resizebox{0.4\columnwidth}{!}{
\includegraphics{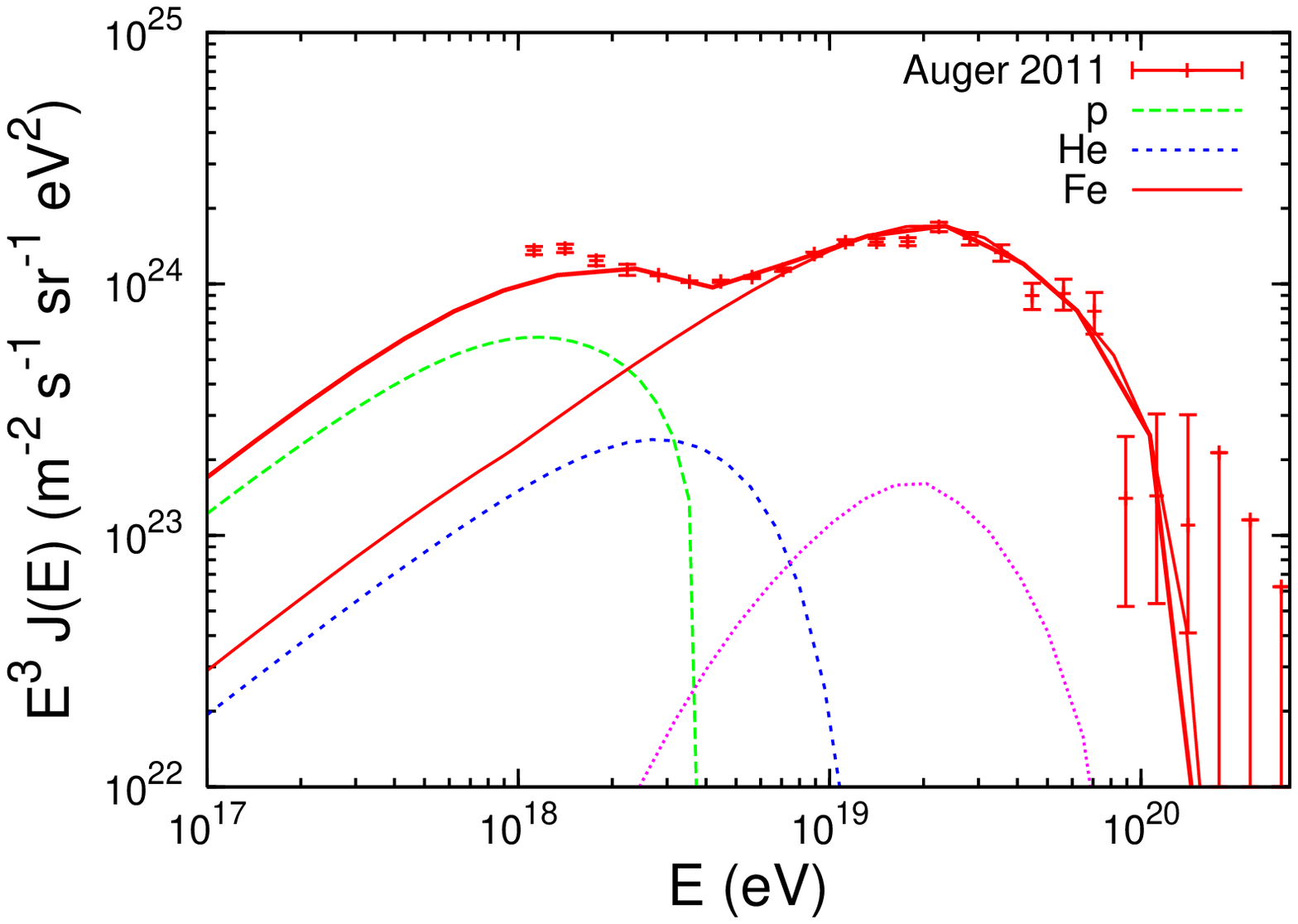} }
\end{tabular}
\caption {\label{fig2}{\small{UHECR flux compared with HiRes and Auger data. Left panel: dip model 
and HiRes data, pure proton injection with a power law index at the source $\gamma_g=2.7$ and a maximum
acceleration energy $E_{max}^p=10^{21}$ eV. Right panel: disappointing model and Auger data, maximum 
proton energy $E_{max}^p=4\times 10^{18}$ eV with a mixture of  p, helium and iron at the source, the 
injection power law index is $\gamma_g=2.0$. 
}}} 
\end{figure}

Concerning sources, in this paper, we will consider the simplest hypothesis of a uniform distribution of sources that
inject UHECR of different species with a power law injection of the type: 
$
Q_{inj}(\Gamma, z) = Q_0(z) \Gamma^{-\gamma_g} e^{-\Gamma/\Gamma_{max}}~,
$
where $\Gamma$ is the Lorentz factor of the particle, $\Gamma_{max}$ is the maximum Lorentz factor that the
sources can provide and $\gamma_g$ is the power law index of the distribution of the accelerated particles at the
sources, being both parameters $\Gamma_{max}$ and $\gamma_g$ independent of the particle type (nucleus or proton). 
The injection of secondary nuclei should be modeled taking into account the characteristics of the 
photo-disintegration process. The dominant process of photo-disintegration is one nucleon ($N$)
emission, namely the process $(A+1) +\gamma_{bkg} \to A+N$, this follows directly 
from the behavior of the photo-disintegration cross-section (see \cite{Nuclei} and 
references therein). Moreover, at the typical energies of UHECR ($E>10^{17}$ eV) 
one can safely neglect the nucleus recoil so that photo-disintegration will conserve 
the Lorentz factor of the particles. Therefore the production rate of secondary 
$A-$nucleus and $A-$associated nucleons will be given by
$
Q_A(\Gamma,z)= Q_p^A(\Gamma,z)=
\frac{n_{A+1}(\Gamma,z)}{\tau_{A+1}(\Gamma,z)}
$
where $\tau_{A+1}$ is the photo-disintegration life-time of the nucleus father $(A+1)$ 
and $n_{A+1}$ is its equilibrium distribution, solution of the kinetic equation for the nucleus $A+1$.

The kinetic equations (\ref{eq:kin_p},\ref{eq:kin_A}) together with the injection of primary and secondary nuclei
can be solved analytically obtaining the equilibrium distribution of primary injected nuclei and all kind of secondaries
produced along the photo-disintegration chain. 

Let us now discuss the implementation of the stochastic treatment of the nuclei propagation as developed in 
the SimProp MC approach \cite{SimProp}. The calculations of the energy evolution and of the photo-disintegration 
life-time (Eq. (\ref{eq:betadisi})) are performed step by step in red-shift. The survival probability as a function of 
red-shift and Lorentz factor of the nucleus $A$ is:
\begin{equation}
P(\Gamma,z) = \exp \left(- \int^{z^{*}}_{z} \frac{1}{\tau_{A}(\Gamma,z^{'})}
\left|\frac{dt}{dz^{'}}\right| dz^{'} \right)  
\label{eq:prob} 
\end{equation}
where $z$ and $z^*$ are the values of the redshift of the current step (from $z^{*}$ to $z$). 
In the standard cosmology the term $|dt/dz|$ is given by$ \left | \frac{dt}{dz} \right |= \frac{1}{(1+z)H(z)} $,
$H(z)=H_0 \sqrt{(1+z)^3 \Omega_m + \Omega_\Lambda}$ with $H_0,\Omega_m,\Omega_\Lambda$ chosen 
according to the standard cosmology \cite{WMAP}.
 
In the SimProp MC code only the photo-disintegration process is treated stochastically while the other processes 
of energy losses are treated as in the kinetic approach, i.e. under the CEL approximation. 

Figure \ref{fig1} shows the good agreement of the fluxes computed with the MC approach and in the kinetic 
approach, for this comparison we have chosen an ideal case of a pure Fe injection at the source with a power 
law injection index $\gamma_g=2.2$. In \cite{SimProp,Nuclei} a more detailed comparison of the results 
obtained with different computation schemes was performed, three different MC schemes were compared 
each other and with the kinetic approach. The fairly good agreement of the fluxes computed with different 
computation schemes shows the solidity of the theoretical approaches used to describe the propagation of 
UHE particles, giving a very useful workbench to test different theoretical models for UHECR sources. 
\begin{figure}[!htb]
\centering
\begin{tabular}{ll}
\resizebox{0.4\columnwidth}{!}{
\includegraphics{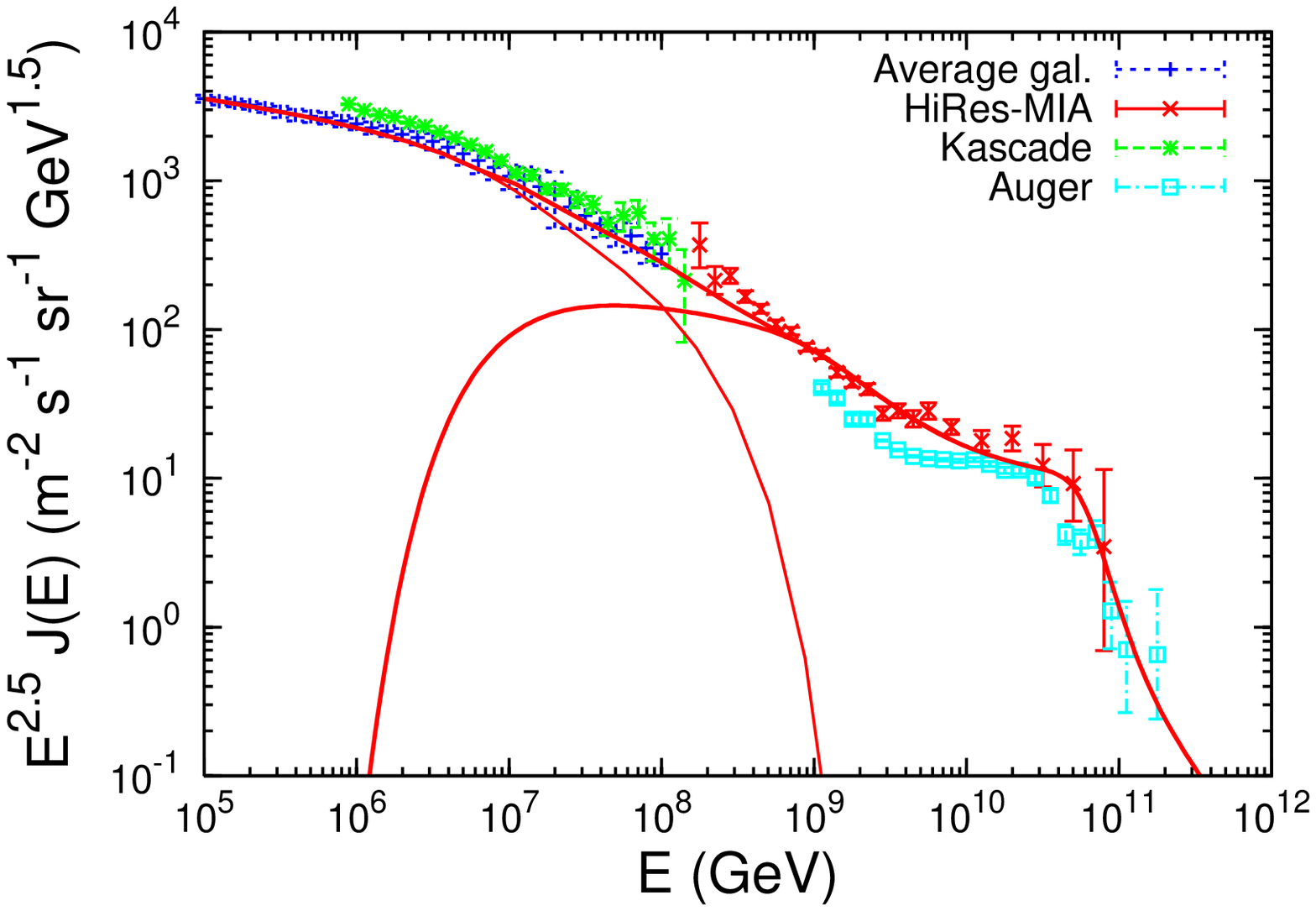} }
&
\resizebox{0.4\columnwidth}{!}{
\includegraphics{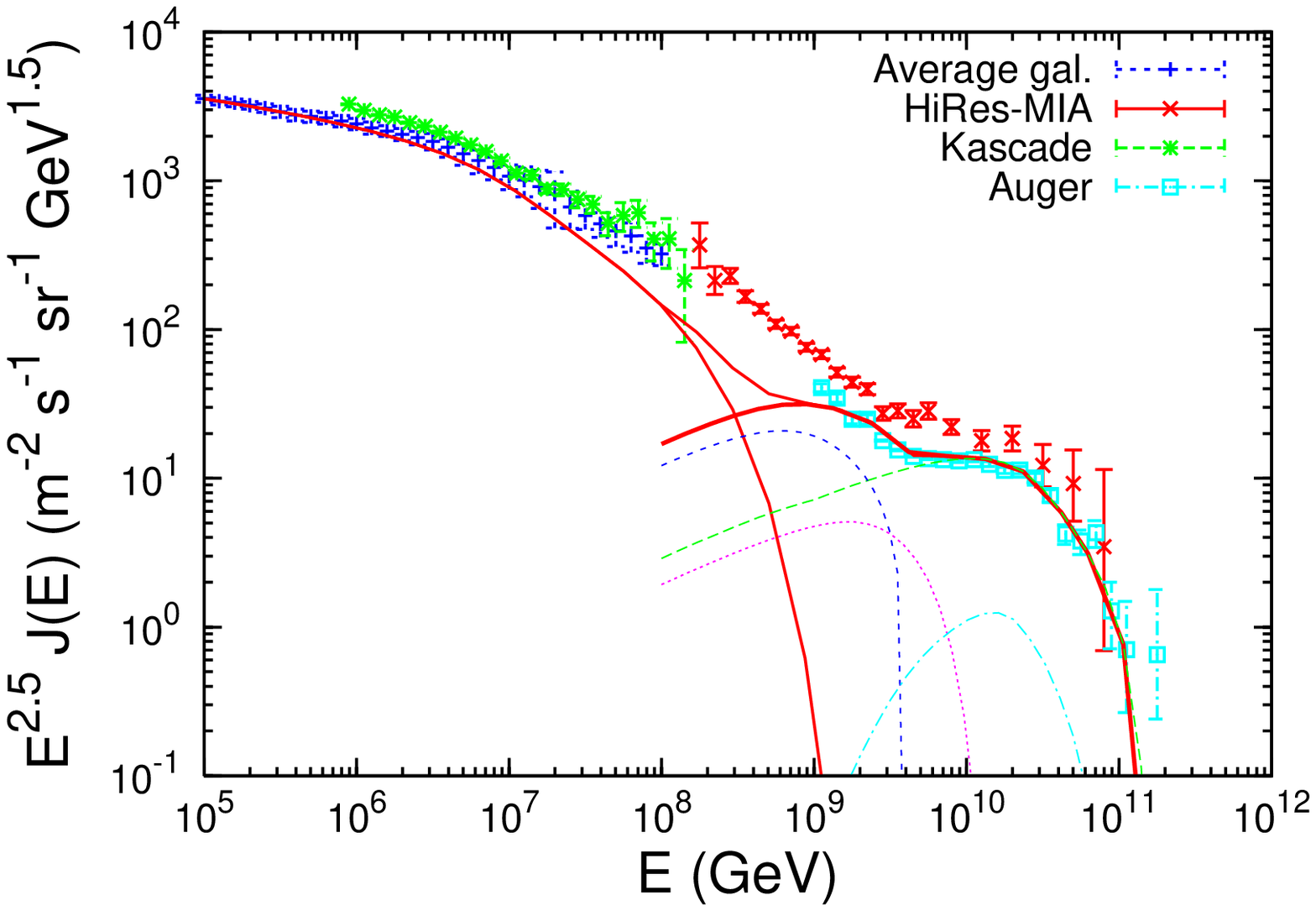} }
\end{tabular}
\caption {\label{fig3}{\small{Transition from galactic to extra-galactic cosmic rays. The galactic cosmic ray flux 
is taken from \cite{AmatoBlasi}. The experimental data (as labeled) are those form HiRes-MIA \cite{Flux}, 
Auger \cite{Flux}, Kaskade \cite{Kascade} and the average over different experimental results as in \cite{averageGal}. 
Fluxes of the extragalactic components are obtained with the same assumptions of figure \ref{fig2}. Left panel: dip model. 
Right panel: disappointing model.
 }}} 
\end{figure}
\section{Transition from Galactic to Extra-Galactic Cosmic Rays}
\label{trans}

In this section we will consider two different models aiming to explain the experimental observations of UHECR. 
In the case of HiRes and TA observations the experimental data are consistent with a pure proton composition,
this experiments favor the dip model that is based on the hypothesis of a pure proton injection at the sources with 
an injection power law index $\gamma_g=2.7$ \cite{dip}. The Auger observations favor a mixed composition at the 
highest energies with a substantial nuclei pollution in the spectrum already at energies $E>4\times 10^{18}$ eV. 
The Auger observations on fluxes can be accommodated in the so-called disappointing model \cite{disapp}, that is 
based on the assumption of a flat injection $\gamma_g=2.0$ and a mixed composition at the sources made of 
protons, helium and iron with an high energy suppression of the fluxes due to the maximum attainable injection energy. 
In the disappointing model primary injected protons are assumed with a relatively low maximum energy 
$E^p_{max} = 4\times 10^{18}$ eV and, in a rigidity dependent scenario, primary injected nuclei will have a maximum 
energy given by $E_{max}^A=ZE_{max}^p$, being $Z$ the atomic mass number of the nucleus $A$. Therefore, in the
disappointing model the behavior of the fluxes at the highest energies is fixed by the maximum energy and it is not a
propagation feature (as the GZK suppression) \cite{disapp}. In figure \ref{fig2} we plot the comparison of the dip model 
with the HiRes data \cite{Flux}  (left panel) and of the disappointing model with the Auger data \cite{Flux} (right panel). 
In the latter comparison we have included in the Fe curve the sum of primary iron and all secondaries with $A\ge40$, 
while the sum of the remaining secondaries with $A<40$ are plotted in the magenta curve below Fe. Figure \ref{fig2} 
shows the agreement of the dip model with the HiRes observations and of the disappointing model with the 
Auger observations. This fact signals once more the puzzling situation we are facing in UHECR physics with 
different experiments favoring alternative scenarios. 

In order to have a deeper view of the problem, trying to asses the intimate nature of UHECR, it is useful to consider 
the transition among galactic and extra-galactic cosmic rays, comparing the matching of the theoretical fluxes with 
experimental data \cite{AloisioBlasi}. In the analysis that follows we will assume a single fixed model for the galactic 
component, namely the flux computed in \cite{AmatoBlasi} taking into account the large scale distribution of supernova 
remnants in the galaxy. In figure \ref{fig3} we plot the total flux with the galactic component of \cite{AmatoBlasi} and the 
extra-galactic components obtained in the two cases of the dip model (left panel) and of the disappointing model (right panel). 
The experimental data of figure \ref{fig3} are those of HiRes-MIA \cite{Flux}, Auger \cite{Flux}, Kaskade 
\cite{Kascade} and the average of the fluxes measured by different experiments as presented in \cite{averageGal}.
The matching of galactic and extragalactic fluxes in the case of the dip model (figure \ref{fig3} left panel) gives a very good 
description of the experimental data, reproducing in an extremely accurate way the spectra observations in the 
intermediate energetic regime where the transition stands. The case of the disappointing model gives a less accurate 
description of the experimental data (figure \ref{fig3} right panel) with a slightly suppression of the theoretical flux in 
the transition region not seen experimentally. 

\section{Conclusions}
\label{conclude}

The theoretical study of the propagation of UHECR through astrophysical backgrounds has reached a remarkable 
level of refinement with several different approaches providing very reliable results. Here, in particular, we 
have compared the kinetic approach of \cite{Nuclei} with the MC approach presented in \cite{SimProp} showing 
the very good agreement of their results (figure \ref{fig1}). Using these theoretical tools it is possible to study with 
good accuracy different models for the production of UHECR by comparing experimental data with theoretical 
expectations. 

In the present paper we have considered the dip model \cite{dip} and the disappointing model \cite{disapp}. 
The dip model is based on the assumption of a pure proton composition of UHECR with an injection power law 
index $\gamma_g>2.5$, the experimental data of HiRes and TA confirm with very good accuracy this model 
(figure \ref{fig2}, left panel) while the Auger data disfavor it with an heavier composition at the highest energies. 
The disappointing model, on the other hand, is based on the assumption of a mixed composition at the sources with 
an injection power law index $\gamma_g<2.5$ and a relatively low maximum energy for protons 
($E_{max}^p=4\times 10^{18}$ eV). This model provides a very accurate description of the Auger data 
(figure \ref{fig2}, right panel), while it does not reproduce the HiRes and TA observations.  
To better characterize these theoretical models, we have also focused the attention on the transition among 
galactic and extra-galactic CR assuming the galactic component as computed in \cite{AmatoBlasi}. In the case 
of the dip model the transition is a sharp change from heavy (galactic) to light (extra-galactic) dominance, starting at 
energies around $10^{17}$ eV. The total flux in the case of the dip model shows a very good agreement with the all 
particle spectrum observed by different experiments (figure \ref{fig3}, left panel). In the case of the disappointing 
model the transition is placed at energies around $10^{18}$ eV and the total flux seems not in perfect agreement 
with observations in particular at energies around the transition region (figure \ref{fig3}, right panel). 

\section*{Aknoweledegements}
I'm grateful to V. Berezinsky, P. Blasi, A. Gazizov and S. Grigorieva with whom several results presented here 
were obtained. I want also to express my gratitude to the Auger group of L'Aquila University for our joint activity 
on the SimProp simulation code.


\begin{thebibliography}{}

\bibitem{Auger}
Auger collaboration, Phys. Letters {\bf B 685} (2010) 239.

\bibitem{HiRes}
HiRes collaboration, Phys. Rev. Lett. {\bf 100} (2008) 101101;
P. Sokolsky,  arXiv:1010.2690 [astro-ph.HE].

\bibitem{TA}
C. Jui {\it et al.} (Telescope Array Collaboration), Proc. APS DPF Meeting arXiv:1110.0133.

\bibitem{Nuclei} 
R. Aloisio, V. Berezinsky and S. Grigorieva, arXiv:0802.4452;
R. Aloisio, V. Berezinsky and S. Grigorieva, arXiv:1006.2484.

\bibitem{Boncioli} 
R. Aloisio and D. Boncioli, Astropart. Phys. {\bf 35} (2011) 152-160.

\bibitem{GZK}
K. Greisen, Phys. Rev. Lett. {\bf 16}, 748 (1966);
G.T. Zatsepin and V.A. Kuzmin, Pisma Zh. Experim. Theor. Phys. {\bf 4}, 114 (1966). 

\bibitem{dip}
V. Berezinsky, A. Gazizov and S. Grigorieva, Phys. Rev. D {\bf 74}, 043005 (2006);
R. Aloisio, V. Berezinsky, P. Blasi, A. Gazizov, S. Grigorieva and B. Hnatyk, Astrop. Phys. {\bf 27}, 
76 (2007).

\bibitem{Chem}
Auger collaboration, Phys. Rev. Lett. {\bf 104} (2010) 091101;
HiRes collaboration, Phys. Rev. Lett. {\bf 104} (2010) 161101.

\bibitem{TransReview}
R. Aloisio, V. Berezinsky and A. Gazizov, "Transition from Galactic to Extra-Galactic Cosmic Rays", review in preparation.

\bibitem{Flux}
R. U. Abbasi et al. [HiRes Collaboration], Phys.Rev.Lett. {\bf 100} (2008) 101101;
F. Salamida for the Pierre Auger Collaboration, Proc. 32th ICRC (Beijing, China) 2011, arXiv:1107.4809.

\bibitem{SimProp}
R. Aloisio, D. Boncioli, A.F. Grillo, S. Petrera and F. Salamida, arXiv:1204.2970.

\bibitem{WMAP}
D.N. Spergel et al. [WMAP collaboration], Astrophys. J. Suppl. {\bf 148} (2003) 175;
D.N. Spergel et al., Astrophys. J. Suppl. {\bf 170} (2007) 377.

\bibitem{disapp} 
R. Aloisio, V. Berezinsky and A. Gozizov, Astrop.Phys. {\bf 34} (2011) 620. 

\bibitem{AloisioBlasi}
R. Aloisio and P. Blasi, in preparation. 

\bibitem{AmatoBlasi}
P. Blasi and E. Amato, JCAP {\bf 1201} (2012) 010. 

\bibitem{Kascade}
W.D. Apel et al. [Kascade-Grande collaboration], Phys.Rev.Lett. {\bf 107} (2011) 171104.

\bibitem{averageGal}
J. R. H\"orandel, Astrop. Phys. {\bf 21} (2004) 241.

\end{thebibliography}
\end{document}